# ON THE MAGNETIC FIELD INSIDE THE SOLAR CIRCLE OF THE GALAXY: ON THE POSSIBILITY OF INVESTIGATION SOME OF CHARACTERISTICS OF THE INTERSTELLAR MEDIUM WITH USING OF PULSARS WITH LARGE FARADAY ROTATION VALUES

R.R. Andreasyan; H.R. Andreasyan; G.M. Paronyan

1. *Introduction.* The study of the magnetic fields of galaxies and, in particular, of our Galaxy has a great importance for explaining many dynamical and active processes taking place in these objects (see, for example, [1-3]). The presence of the magnetic field of the Galaxy can explain the transportation of cosmic rays through the interstellar medium, as well as synchrotron background radiation in the Galaxy [4,5]. The magnetic field of the Galaxy was studied using observational data of various types, such as interstellar polarization of starlight, Zeeman splitting of spectral lines of HI and different molecules in the radio range, data on Faraday rotation of extragalactic radio sources and pulsars (see, for example, [6-8]). It is known that pulsars, for which numerous and diverse observational data were obtained, can be considered probes for studying the interstellar medium. In particular, data on dispersion measures (DM), which practically are known for all known pulsars, and about measures of Faraday rotation (RM) (of the order of 1150 pulsars) are very important for studying the magnetic field of the Galaxy. These data are directly derived from observations of pulsars. Theoretically they are expressed by the electron density $n_e$ in the interstellar medium through which the polarized radio emission of the pulsar passes and the projection of the magnetic field $B_L$ (in Gauss) in this medium, using the following formulas:

$$DM = \int n_e dL , \qquad (1)$$

$$RM = \alpha \int n_e B_L dL , \quad (\alpha = 8.1 \cdot 10^5) . \qquad (2)$$

In these formulas, integration is carried out over the entire traversed path of radiation (L in parsecs) from the pulsar to the observer. Formula 1 makes it possible to determine the distance of a pulsar with the known electron density distribution in the Galaxy, and formula 2 together with formula 1 makes possible to determine the average component of the tension of interstellar magnetic field $[B_L]$ on the line of sight in microgauss (μG).

$$[B_L] = (1/\alpha)(RM)/(DM) = 1.23 \, (RM)/(DM). \qquad (3)$$

Data RM and DM were used to study the structure and magnitude of the magnetic field of the Galaxy, since the seventies of the last century, when the rotation measures were known for only 3-4 tens of pulsars [9, 10]. As the amount of RM data increases, more detailed studies have been carried out and various models have been proposed for the plane component of the Galactic magnetic field [11-14], as well as for the magnetic field in the Halo of the Galaxy (see, for example, [15-17]). In particular, in the work of Andreasyan and Makarov [15], where the model of the two-component magnetic field of our Galaxy was proposed, for the first time it was shown that the data on the rotation measures of pulsars and extragalactic radio sources are in good agreement with the model when the magnetic field of the plane component of the spiral arms is imbedded in the magnetic field of the galactic halo with the dipole configuration, which is deformed due to the differential rotation of the Galaxy. In [2], using the data of RM, DM and the distance of pulsars, we constructed a two-color map of the plane component of the magnetic field of the Galaxy, on which large-scale regions with regular magnetic fields, approximately corresponding to the spiral structure of the Galaxy, were clearly distinguished.

It should be noted that in all the works mentioned above, three models of the plane component of the magnetic field of the Galaxy are mainly discussed: 1) a bisymmetric spiral (BSS), in which the magnetic field in neighbor spiral arms has opposite directions; 2) an axially symmetric (ASS) structure with two changes in the direction of the field inside the solar circle; 3) concentric circular model. In particular, it was shown in [14] that in the general galactocentric magnetic field, which is directed clockwise, the magnetic field in the ring with a galactocentric distance of 5-7 kpc directed counterclockwise is mainly emitted. There were also proposed models in which the spiral structure of the magnetic field roughly coincides with the inter arm regions of the Galaxy. There are also works (see, for example, [18]) in which it is shown that none of these models correspond to observational data better than the other. It should be noted, that in addition to the spiral arms inside the solar circle, magnetic fields outside this circle were also discussed, for example, in the Perseus spiral arm and in the local Orion arm, which is probably a branch from the Perseus arm. The magnetic fields in these arms (see, for example, [19]) have a direction opposite to the inner spiral arm of Sagittarius, in which a fairly regular magnetic field is directed toward the observer.

Note that inside the solar circle, in addition to the spiral arm of Sagittarius, there are two other spiral arms (see, for example, [17, 20, 21]), whose magnetic fields are also studied using the data of RM of pulsars located inside or further of these arms see Fig. 1).

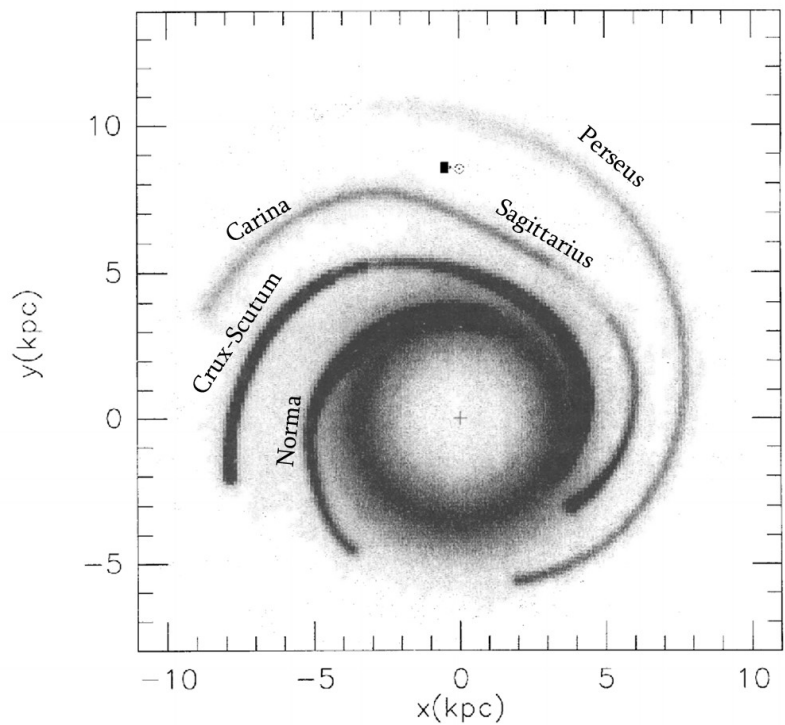

Fig.1. Distribution of electron concentration $n_e$ in the Galactic plane. In the vicinity of the Sun $n_e \approx 0.019$ cm$^{-3}$. The largest electron concentration corresponds to the darkest part in the figure.

Figure 1 shows the distribution of the electron density $n_e$ in the Galactic plane, taken from [20]. As follows from Fig. 1, the polarized radiation of distant pulsars observed by us passes through several spiral arms and carries total information on the magnetic field of these arms. These data are used to construct the above-mentioned models of the plane component of the Galactic magnetic field. If we take into account the fact that the error in determining the distances of distant pulsars in directions closer to the center of the Galaxy due to this model of electron concentration can reach tens of percents [20], as well as the presence of small-scale fluctuations of the magnetic field, the intensity of which reaches (or even exceeds) the same values as in regular fields, it becomes clear that the task of choosing a more plausible model of the magnetic field from the above is very difficult or even impossible at the present time.

In this paper, to study the magnetic field in regions inside the solar circle, we will use pulsars with large rotation measures $|RM| > 300 \, rad/m^2$, closer to the center of the Galaxy. It is clear that: firstly, distant pulsars basically have large rotation measures, and, second, the numerous data of pulsars with small values $|RM|$ and with significant errors in the distances, can clutter the graphs very much, from which important details of the distribution of the magnetic field can be lost. The limitation $|RM| > 300 \, rad/m^2$ is also due to the fact that, as will be seen later, these pulsars are mainly concentrated in the region around the galactic center in Galactic longitudes $0^0 < l < 90^0$, and $270^0 < l < 360^0$ (in directions to the anti-center of the Galaxy we have only 6 this kind of pulsars), and, consequently, they can be used to study the magnetic field in regions inside the solar circle of the Galaxy.

*2. Distribution of pulsars along galactic longitude.* The data from the ATNF pulsar catalog are used in the work [22]. In this paper we use data from 199 pulsars with $|RM| > 300 \, rad/m^2$, selected from 1133 pulsars with known rotation measures. It turns out that pulsars with large values of $|RM|$ basically (with the exception of 10 pulsars) are concentrated near the plane of the Galaxy in a layer of ± 500 parsecs. Figure 2 shows the distribution of Faraday rotation measures over galactic longitude l. In constructing the plot, almost all pulsars with $|RM| > 300 \, rad/m^2$ were used. For the compactness of the figure, 6 pulsars with $|RM| > 2000 \, rad/m^2$, as well as 6 pulsars that are in the directions of the galactic longitude $90^0 > l > 270^0$ (in the direction of the anticenter) were excluded from consideration. In the figure, the galactic longitudes $-90^0 < l < 0^0$ correspond to the coordinates - $(360^0 - l)$.

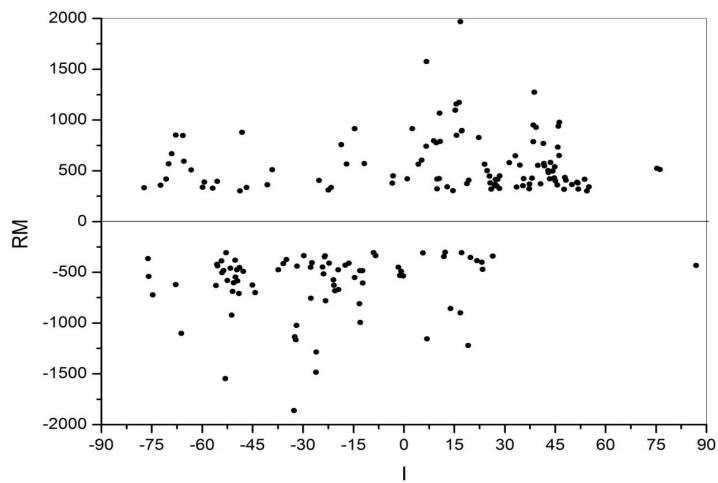

Fig.2. Distribution of Faraday rotation measures $|RM| > 300 \, rad/m^2$ over galactic longitude l.

It can be seen from Fig. 2 that the magnitudes of the rotation measures with respect to the galactic coordinate's l are distributed rather chaotically. Sometimes pulsars with angular distances less than one or two degrees have rotation measures that differ even in sign. There are more than twenty such examples in the list of used pulsars and on the graph. Therefore, we can assume that the characteristic size of the chaotic distribution of the rotation measures corresponds to angular distances in the order of 1-2 degrees. In many spatially close pulsars, even if the Faraday rotation signs coincide, their magnitudes sometimes differ by several times. This can mean that such rotation measures can not be formed in the large-scale magnetic field of the Galaxy.

We note that when analyzing the map of the magnetic field of our Galaxy [23], we showed that one of the largest formations with a fairly regular magnetic field in the Galaxy is the spiral arm of Sagittarius. But even in this direction the magnetic field does not reach such large values. A regular magnetic field in the direction of Sagittarius's spiral arm is also seen in Fig. 2 in galactic longitudes $30^0$-$70^0$.

From the foregoing, it can be assumed that a very large rotation measure of pulsar is probably due to the influence of one nearby object (relative to the pulsar), which has large electron concentrations $n_e$, and perhaps also a large value of the average magnetic field strength $B_L$. Such objects are projected on pulsars, and the polarized radiation of pulsars passing through them acquires a large Faraday rotation. In this connection we note that in our early work [24] it was shown that large rotation measures |RM| unidentified extragalactic radio sources located in directions close to the plane of our Galaxy can also be explained by the passage of polarized radio emission of these radio sources through the HII regions located in the Galactic plane. The conclusion that pulsar radiation can pass through nearby objects with large electron concentrations $n_e$ can also be based on certain features of the distribution of pulsars with large rotation measures. For example, the majority of pulsars with large rotation measures are mainly located near or inside a 4-kiloparsec ring with a large electron concentration [20]. It can be assumed that this ring is patchy and consists of numerous condensations with dimensions of about 100 parsecs or less. At distances of the majority of the used pulsars, such dimensions correspond to one or two angular degrees, that is, of the order of the characteristic distance of the chaotic distribution of the rotation measures (see Fig. 2). Passing through such condensations, the polarized radiation of pulsars can undergo additional Faraday rotation of different signs and different magnitudes, depending on the magnetic field in these objects. However, in most cases it is most likely that this additional Faraday rotation is due to the effect of only one condensation. Such an assumption can be justified by the following arguments: from the analysis of rotation measures (RM) and dispersion measures (DM) of pulsars with large Faraday rotation values, it turns out that the magnetic field averaged over the entire path is 1-3 micro gauss, as in other sections in the interstellar medium (see formulas 1, 2 and 3, and also Fig. 3).

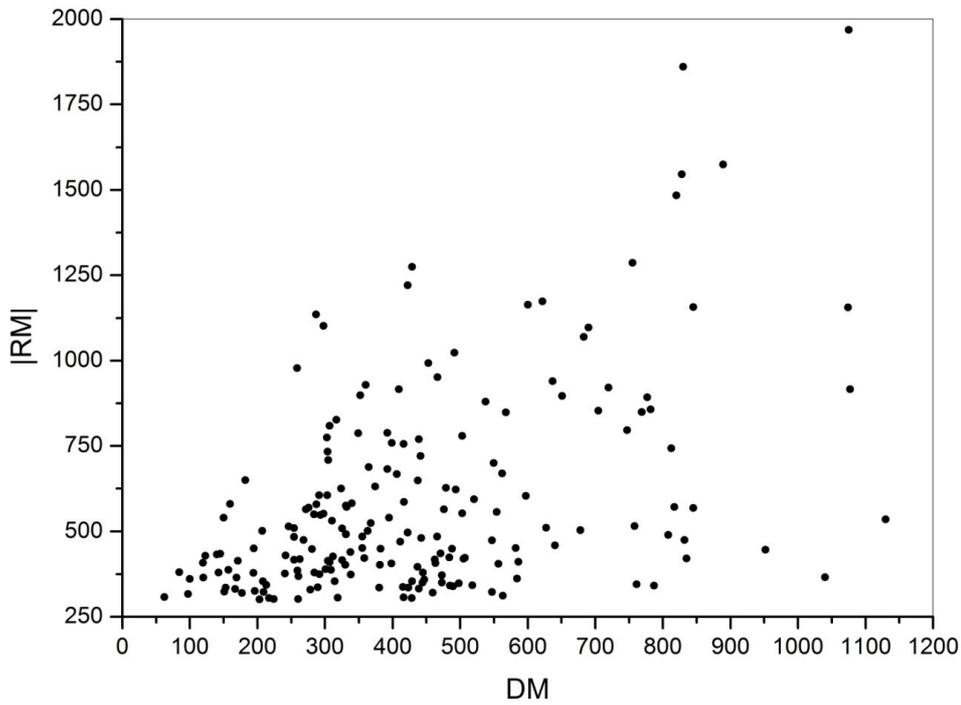

Fig.3. The dependence |RM| from DM for pulsars with |RM|> 300rad / m$^2$

From formula 2 it follows that the dispersion measure is additive, and increases when the pulsar radiation passes through each new condensation. However, the rotation measure does not have such an additive property (see formula 1), because, in addition to the electron concentration, it also depends on the direction and magnitude of the magnetic field. Assuming that both the rotation measure and the dispersion measure are formed due to passage of the pulsar radiation through several condensations, the dispersion measure DM would increase after each passage in a linear manner, while the rotation measure RM would decrease due to compensation of the Faraday rotation by separate condensations with different directions of the magnetic field, or even increased, it will be more slowly. This is due to the fact that it is difficult to imagine that several clouds with the same direction of the magnetic field are located on the line of sight, while a few clouds are also located in the very close direction, but with the opposite direction of the magnetic field. It follows from the above that when a polarized radio emission of a pulsar passes through several clouds, there should not be a noticeable positive correlation between |RM| and DM, while in Figure 4 there is a noticeable correlation between them. Although this correlation is weak (the correlation coefficient is in the order of 0.5), this too can be considered as indirect evidence in favor of the assumption that in most cases the additional Faraday rotation is due to the influence of one condensation.

*3. Distribution of pulsars with large RMs in the Galactic plane*. In the previous section it was suggested that the large Faraday rotations of pulsars are in most cases due to the contribution of one cloud (presumably HII regions, molecular clouds or dark nebulae) located between the pulsar and the observer. The magnetic field in this cloud can be oriented along the direction of a large-scale field of the Galaxy, or simply a deformed continuation of the galactic field. Consequently, data on pulsars with large |RM| can contain information about the galactic magnetic field and can be useful for studying the large-scale magnetic field of the Galaxy. Figure 4 shows the distribution of pulsars with |RM|> 300rad / m$^2$ on the plane of the Galaxy. The coordinate axes pass through the center of the Galaxy. The coordinates of the Sun, indicated by

an asterisk - (0 kpc; 8.5 kpc). Pulsars are indicated by circles. The black circles correspond to pulsars with positive values of RM (the projection of the magnetic field is directed toward the observer), and the white circles correspond to pulsars with negative values of RM. The figure clearly identifies the ring located between galactocentric circles with radii R about 5 and 7 kilo parsec. In this ring, almost all pulsars on the right side (with the exception of 3 pulsars) have Faraday rotations with a positive sign, and on the left side, except for 4 pulsars, all the others have a RM with a negative sign. This distribution of the signs of the pulsars RM in an excellent manner corresponds to the direction of the magnetic field counterclockwise in the ring with 5kpc<R<7 kpc (case of 7 pulsars, which are an exception to this scheme, can be investigated separately). In the remaining areas, it is difficult to highlight a sufficiently large-scale region with such a clearly marked direction of the magnetic field. This probably can be a consequence of the fact that in the direction of the center of the Galaxy, where the line of sight passes through several galactic arms, the error in determining the distances of pulsars is larger.

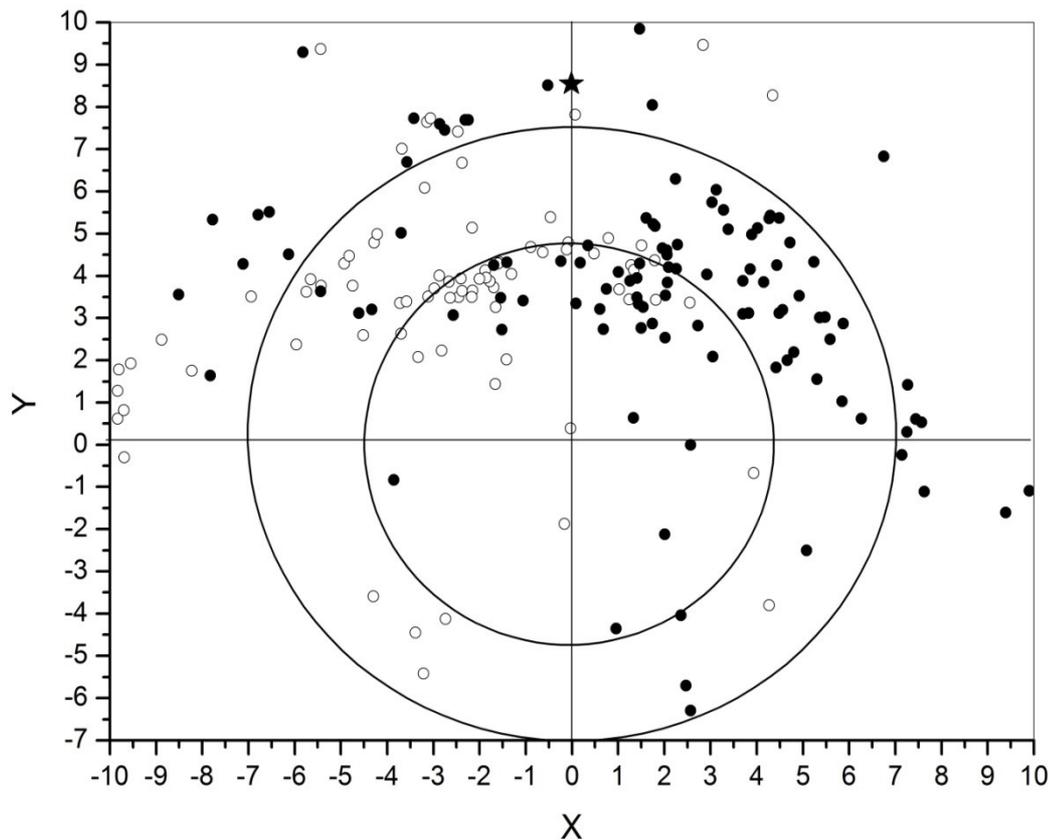

Fig.4. Distribution of rotation measures of pulsars in the plane of the Galaxy. Black circles denote pulsars in RM> 300rad / m² (the projection of the magnetic field on the line of sight is directed toward the observer), white circles - in which RM < -300rad / m² (the projection of the magnetic field is directed from the observer).

Thus, the analysis of pulsar data with |RM|> 300rad / m² partially corresponds to the model of the magnetic field of the Galaxy proposed in [14], which proposes a circular model of the magnetic field of the Galaxy, directed clockwise, with a unique change in the direction of the field in the galactocentric ring with 5 kpc <R <7 kpc, where the magnetic field is directed counter-clockwise. Once again, we recall that we were supposed that large |RM| are due to the contribution of regions projected on the pulsar with an increased electron concentration in which

the magnetic field is oriented along the direction of the large-scale field of the Galaxy, or simply is a deformed extension of the Galactic field.

*4. Identification of pulsars with large RM*. As was suggested in the previous sections, large rotation measures of pulsars can be caused by clouds with an increased concentration of free electrons through which the polarized radio emission of pulsars passes. To determine what these clouds represent, we searched for interesting objects in the region of 5 angular minutes around pulsars with large RM-s. To search for objects, we used the database "SIMBAD". At distances of the considered pulsars, the 5-minute region is a circle with a diameter of the order of 10-15 parsecs. So, if these objects are closer to us than the pulsar and fall into this region, then they probably have dimensions of the order of 10-15 parsecs, or less. Thus, searches were carried out around 199 pulsars with the $|RM|> 300$ rad / m$^2$. As a result, around 70 pulsars, nebular objects were found, mainly: HII regions ~ 35%, dark nebulae ~ 45%, and molecular clouds ~ 20%.

It is known that the characteristic sizes of HII regions and dark nebulae are larger or of the order of 10-15 parsecs, and since the electron density $n_e$ is much larger in them than in the interstellar medium (on average 0.03 cm$^{-3}$) and the magnetic field strength is greater or in the order of the mean field of interstellar medium (2-3 micro gauss), then indeed part of the large rotational measures of the mentioned 70 pulsars can be explained by the passage of polarized radio emission of pulsars through HII regions and dark nebulae. Simple estimates show that the contribution to the RM of the HII region or the Stremgren zones around young O-B stars which radius can reach about 100 parsecs, the electron concentration $n_e$ is in the order of 1 cm$^{-3}$, and the magnetic field, as in the interstellar medium, is in the order of 2-3 micro gauss, can be in the order of 320-480 rad / m$^2$. Similar estimates can be obtained for dark nebulae whose dimensions are smaller than the size of the HII regions, but the electron concentration is probably much larger than in the HII regions. This means that the contribution of HII regions, molecular clouds and dark nebulae to the large value of the Faraday rotation of the pulsar can be decisive.

*5. Conclusion*. In conclusion, we note that using the observational data of pulsars with large Faraday rotation values, several results have been obtained in this paper. First, it was shown that the galactic distribution of the pulsars with rotation measures $|RM|> 300$ rad / m$^2$ better corresponds to the circular model of the magnetic field of the Galaxy [14], with the counterclockwise direction of the magnetic field in the galactocentric circle 5 kpc <R <7 kpc. It was also suggested and justified that large $|RM|$ may be due to the contribution of regions with an increased electron concentration (probably HII regions, dark nebulae and molecular clouds), projected on the pulsar. In these objects, the magnetic field can be oriented along the direction of a large-scale field of the Galaxy, or simply be a deformed extension of the galactic field.

Note that in order to study the large-scale magnetic field of the Galaxy, data from more than 150 HII regions and molecular clouds in which the magnetic fields were estimated by different methods (mainly by the Zeeman splitting of spectral radio lines) were also used in [25]. It was also suggested that the magnetic fields in these objects can be a relic of the magnetic field of the Galaxy. We made an attempt to identify these objects with ours, but the direction of none of them coincided with the coordinates of the pulsars in this paper.

It should also be noted that from the conclusions of this paper a new possibility arises for estimating magnetic fields and some other characteristics of those HII regions, molecular clouds and dark nebulae that are closer than the pulsar, but are projected on it. This can be done by estimating the contribution of these nebular objects in the Faraday rotation of this pulsar.

The authors are grateful to the authors of the pulsar catalog (ATNF Pulsar Catalog), http://www.atnf.csiro.au/research/pulsar/psrcat/; Manchester, R. N., Hobbs, G.B., Teoh, A. & Hobbs, M., AJ, 129, 1993-2006 (2005).

Byurakan Astrophysical Observatory named after V.A. Ambartsumyan
e-mail: randrasy@bao.sci.am; handreasyan@bao.sci.am; gurgen@bao.sci.am;

ON THE MAGNETIC FIELD INSIDE THE SOLAR CIRCLE OF THE GALAXY
ON THE POSSIBILITY OF INVESTIGATION SOME OF CHARACTERISTICS OF THE INTERSTELLAR MEDIUM WITH USING OF PULSARS WITH LARGE FARADAY ROTATION VALUES

R.R. Andreasyan; H.R. Andreasyan; G.M. Paronyan

To study some characteristics of the interstellar medium, observational data of pulsars with large Faraday rotation values ($|RM| > 300$ rad / m$^2$) were used. It was suggested and justified that large $|RM|$ values can be due to the contribution of the regions with increased electron concentration, projected on the pulsar. Most likely these are the HII regions, dark nebulae and molecular clouds. In these objects the magnetic field can be oriented in the direction of a large-scale field of the Galaxy, or simply is a deformed extension of the galactic field. It was shown that the Galactic distribution of rotation measures of pulsars with $|RM| > 300$ rad / m$^2$ corresponds to the circular model of the magnetic field of the Galaxy, with the counter-clockwise direction of the magnetic field in the galactocentric circle 5 kpc $< R <$ 7 kpc.